\documentclass[conference]{IEEEtran}
\IEEEoverridecommandlockouts
\usepackage{algorithm,algorithmic,amsbsy,amsmath,amssymb,epsfig,subfig,bbm,mathrsfs,mathdots,arydshln,arydshln,multirow,verbatim,booktabs,setspace,url,stfloats,fixltx2e,cite,nomencl,epstopdf}

\usepackage{xcolor}
\def\BibTeX{{\rm B\kern-.05em{\sc i\kern-.025em b}\kern-.08em
    T\kern-.1667em\lower.7ex\hbox{E}\kern-.125emX}}

\newcommand{\beq}{\begin{equation}}
\newcommand{\eeq}{\end{equation}}
\newcommand{\bitm}{\begin{itemize}}
\newcommand{\ba}{\begin{array}}
\newcommand{\ea}{\end{array}}
\newcommand{\eitm}{\end{itemize}}
\newcommand{\beqn}{\begin{eqnarray}}
\newcommand{\eeqn}{\end{eqnarray}}
\newcommand{\beqno}{\begin{eqnarray*}}
\newcommand{\eeqno}{\end{eqnarray*}}
\newcommand{\bma}{\begin{displaymath}}
\newcommand{\ema}{\end{displaymath}}
\newcommand{\bnu}{\begin{enumerate}}
\newcommand{\enu}{\end{enumerate}}
\newcommand{\bce}{\begin{center}}
\newcommand{\ece}{\end{center}}
\newcommand{\btb}{\begin{tabular}}
\newcommand{\etb}{\end{tabular}}

\begin{document}
%\title{Resource Allocation in Quantum-Secured Space-Air-Ground Integrated Networks}
\title{Resource Allocation in Quantum Key Distribution (QKD) for Space-Air-Ground Integrated Networks}
\author{\IEEEauthorblockN{Rakpong Kaewpuang$^{\mathrm{1}}$, Minrui Xu$^{\mathrm{1}}$, Dusit Niyato$^{\mathrm{1}}$, Han Yu$^{\mathrm{1}}$, and Zehui Xiong$^{\mathrm{2}}$} \\
\IEEEauthorblockA{ $^{\mathrm{1}}$School of Computer Science and Engineering, Nanyang Technological University (NTU) \\
				    $^{\mathrm{2}}$Pillar of Information Systems Technology and Design, Singapore University of Technology and Design (SUTD)       
       }}
\maketitle

\begin{abstract}

Space-air-ground integrated networks (SAGIN) are one of the most promising advanced paradigms in the sixth generation (6G) communication. SAGIN can support high data rates, low latency, and seamless network coverage for interconnected applications and services. However, communications in SAGIN are facing tremendous security threats from the ever-increasing capacity of quantum computers. Fortunately, quantum key distribution (QKD) for establishing secure communications in SAGIN, i.e., QKD over SAGIN, can provide information-theoretic security. To minimize the QKD deployment cost in SAGIN with heterogeneous nodes, in this paper, we propose a resource allocation scheme for QKD over SAGIN using stochastic programming. The proposed scheme is formulated via two-stage stochastic programming (SP), while considering uncertainties such as security requirements and weather conditions. Under extensive experiments, the results clearly show that the proposed scheme can achieve the optimal deployment cost under various security requirements and unpredictable weather conditions.

\end{abstract}


\section{Introduction}
\label{sec:introduction}

%basic concept of QKD
Quantum key distribution (QKD) has recently been introduced as a secure secret key distribution solution to achieve information-theoretic security (ITS)~\cite{y-cao-the-evolution2022}. The utilization of QKD over space-air-ground integrated networks (SAGIN), i.e., QKD over SAGIN, will be promising since QKD can protect communications among heterogeneous nodes from security threats, such as man-in-the-middle and eavesdropping attacks. This unique advantage comes from the applications of the solid principles of quantum physics, such as the quantum no-cloning theorem and the Heisenberg's uncertainty principle \cite{m-mehic-quantum-key-distribution2020}. In QKD over SAGIN, the QKD can provision secret keys by means of quantum channels to establish secure communication channels among space, aerial, and ground nodes.

% protect an information from attacks, i.e., man-in-the-middle and eavesdropping attacks, since the QKD takes advantage of the principles of quantum physics to distribute secure secret keys among QKD nodes \cite{y-cao-the-evolution2022}. By exploiting the quantum no-cloning theorem and the Heisenberg's uncertainty principle in the principles of quantum physics \cite{m-mehic-quantum-key-distribution2020}, the QKD can provision secret keys by means of quantum channels to achieve the information security which is indestructible. The quantum channels are used for quantum transmission that information is conveyed by quantum states (i.e., photon polarization states).

\subsection{Related Works}
\label{sec:related-works}
    
%\cite{m-mehic-a-novel2019}

%QKD-SAGIN
The QKD over SAGIN was proposed to achieve secure communications in a global-coverage and reconfigurable manner \cite{h-cui-space-air-ground2022}, \cite{m-xu-quantum-secured-space2022}. The authors \cite{m-xu-quantum-secured-space2022} proposed the QKD service provisioning framework to establish secure communications in quantum-secured SAGIN. The two-stage SP was formulated to minimize the provisioning cost under the uncertainty of secret-key rates. In practice, the framework was applied to support Metaverse applications to achieve the optimal solution. In \cite{y-cao2021-hybrid}, the network architecture of hybrid trusted/untrusted relay-based QKD over optical backbone networks was proposed. In their architecture, the overall deployment cost of ground QKD network components was optimized by the integer linear programming model and the heuristic algorithm. The authors \cite{kaewpuang2022-adaptive} proposed the hierarchical architecture for QKD-secured federated learning (FL) systems. In \cite{kaewpuang2022-adaptive}, the QKD resource management was formulated, which is based on the two-stage SP, to minimize the deployment cost under the uncertainty of the secret-key rate requirements from FL applications. \cite{s-k-liao-satellite-to-ground2017}, \cite{j-yin-entanglement-based2020} introduced systems to provide QKD services via free space to secure applications to satisfy security requirements. The authors in \cite{j-yin-entanglement-based2020} demonstrated, the quantum satellite (i.e., Micius) that successfully distributes the secret key (i.e., with the rate of 0.12 bits per second) between two ground nodes via free space. Drone-based entanglement distribution was proposed in \cite{h-y-liu-optical-relayed2021}, \cite{h-y-liu-drone-based-entanglement2020}. In \cite{h-y-liu-drone-based-entanglement2020}, the mobile quantum communications via entanglement distribution were demonstrated and the proposed system can tolerate various weather conditions (i.e., daytime, clear nights, and rainy nights). By comparison with fiber-based and satellite quantum communication, drone-based entanglement distribution has the advantages of unparalleled mobility, flexibility, and reconfigurability. 

However, none of these existing works considers the problem of optimizing the QKD resource allocation for QKD-secured SAGIN. Moreover, current works largely overlook the uncertainties of secret-key rates and weather conditions affecting the distribution of quantum secret keys.  

\subsection{Solution Approach and Contribution}

In QKD over SAGIN, two remote QKD nodes can exchange secret keys by encoding them into quantum bits (qubits) and transmitting them via optical fibers or free space \cite{y-cao-the-evolution2022}. The optical fiber-based QKD~\cite{y-cao2021-hybrid} option is used to transmit qubits and bits in a single fiber by leveraging the wavelength-division multiplexing (WDM) technique. However, qubits are not readily amplified due to the quantum no-cloning theorem, and thus the secret-key rates are exponentially decreased when the distance between two QKD nodes increases. Alternatively, the UAV-based QKD~\cite{h-y-liu-drone-based-entanglement2020} and the satellite-based QKD~\cite{s-k-liao-satellite-to-ground2017} can transmit qubits via free space, which are mobile and more flexible to deploy. Moreover, the unmanned aerial vehicle (UAV)-based QKD and satellite-based QKD are supplementary options to alleviate the limitation in the distance of the optical fiber-based QKD. Therefore, the QKD over SAGIN consists of optical fiber-based, UAV-based, and satellite-based QKD is an efficient solution for quantum-secured SAGIN for secure global communications. However, due to the lack of efficient resource allocation schemes for QKD resources in quantum-secured SAGIN, this system may not be ready for practical deployment in the real world. In this paper, we propose a resource allocation scheme based on stochastic programming to optimize the deployment cost of QKD over SAGIN. In the proposed scheme, the optimal allocation solution is obtained under various uncertainties of security requirements and weather conditions that affect the utilization of QKD over SAGIN. The major contributions of this paper can be summarized as follows:
% However, due to a lack of efficient resource allocation schemes for QKD resources in quantum-secured SAGIN, in this paper, the resource allocation scheme to provision the QKD services to QKD nodes with uncertainties of security requirements and weather conditions that effect on transmission of satellite are introduced. The major contributions of this paper can be summarized as follows:
\begin{itemize}
    \item We propose a novel resource allocation scheme for quantum-secured SAGIN where QKD services can protect secure communications between space, aerial, and ground nodes by exchanging secret keys in quantum channels.
    \item In the proposed scheme, we formulate and solve the SP to obtain the optimal solution on resource allocation (i.e., QKD and KM wavelengths) and routing. Specifically, the QKD resource and routing decisions are jointly obtained under the uncertainties of secret-key rates and weather conditions.      
    \item We show the performance of the proposed model in minimizing deployment costs under various settings.
\end{itemize}

\section{System Model}
\label{sec:system-model}

\begin{figure}[!]
\centering
\includegraphics[width=2.2in]{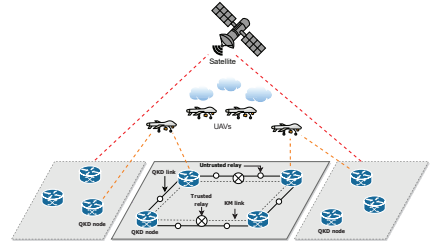}
%\vspace{-0.2cm}
\caption{The system model of quantum-secured SAGIN.}
\label{fig:system-model}
\vspace{-0.5cm}
\end{figure}

We consider a three-layer QKD service provisioning framework for QKD over SAGIN, as shown in Fig.~\ref{fig:system-model}. The three-layer QKD service provisioning framework consists of optical fiber-based QKD services, UAV-based QKD services, and satellite-based QKD services.

For optical fiber-based QKD services, we consider that the QKD nodes are co-located with the backbone nodes in the conventional passive optical networks (PON). There are three types of nodes (i.e., QKD nodes, trusted relays, and untrusted relays) and two types of links (i.e., key management (KM) links and QKD links). As QKD nodes co-located with backbone nodes in PON, the links (i.e., QKD, KM, and optical links) are multiplexed within a single fiber by MUX/DEMUX components \cite{y-cao2021-hybrid}. The secret keys are generated in QKD nodes. The trusted and the untrusted relays are located between QKD nodes. Each QKD node comprises a global key server (GKS), a local key manager (LKM), a security infrastructure (SI), and trusted/untrusted relays. A trusted relay consists of two or multiple transmitters of MDI-QKD (MDI-Txs), LKM, and SI. An untrusted relay comprises one or multiple receivers of MDI-QKD (MDI-Rxs). The MDI-QRx must be located between two MDI-QTxs to exchange local secret keys. The LKM is used to get and store the local secret keys from the connected MDI-QTxs. The LKM performs a secret-key relay via one-time pad method \cite{y-cao2021-hybrid} to establish global secret keys between QKD nodes. The SI is applied to ensure that the trusted relay is in a safe position. For the links, the QKD links are used to connect between MDI-QTxs and MDI-QRxs. The KM links are used to connect between LKMs. Each QKD link contains quantum and classical channels. Each KM link contains a classical communication channel, which is implemented by wavelengths. 

To satisfy all security requests of the ground nodes, UAV- and satellite-based QKD services shall be introduced. For UAV-based QKD services, the trusted UAVs with quantum equipment can provide reconfigurable QKD services to serve ground QKD nodes. For satellite-based QKD services, the satellite-based QKD can provide QKD services to QKD nodes from lower earth orbit (LEO) satellites via free space \cite{h-y-liu-drone-based-entanglement2020}. The satellite-based QKD services can distribute quantum secret keys to ground nodes via free space. However, the transmission of a satellite has to take weather conditions into account since the satellite cannot provide the QKD services if the weather is in a bad condition (e.g., cloudy and rainy). To establish secure communications in quantum-secured SAGIN, two QKD nodes can transmit secret keys encoded as qubits by using QKD links. The QKD nodes receive quantum secret keys from optical fiber-based QKD services when they are connected via optical fibers. However, when the QKD nodes are not connected by optical fibers, they can request the QKD services from UAVs or satellites via free space.    

% system model %

\subsection{The Network Model and The Cost Model}
\label{subsec:network_model}

In the network model, let $\mathcal{F}$ and $f(s_{f}, d_{f}, P_{f}(\cdot)) \in \mathcal{F}$ denote the set of QKD requests and a QKD request of QKD nodes, respectively. $s_{f}$ and $d_{f}$ are the source node and destination node of request $f$, respectively. Let $K_{\Theta}$ denote the maximum achievable secret-key rate at a distance $\Theta$ of a medium (i.e., a fiber link, UAV, and satellite). The number of parallel QKD links $P_{f}(\cdot)$ to satisfy the security demand (secret-key rate) of model transmission between $s_{f}$ and $d_{f}$ can be expressed as $P_{f}(\tilde{\omega}) = \Big\lceil \frac{\tilde{k}_{f}}{K_{\Theta}} \Big\rceil$, where $f$ is the request and $\tilde{k}_{f}$ is a random variable of secret-key rate requirement of request $f$. $\Theta$ is the distance between two connected MDI-QTxs, which can be expressed as $\Theta \approx 2 \cdot \vartheta$, where $\vartheta$ is the distance between the MDI-QRx connecting to the MDI-QTx.  

% \subsection{The Cost Model}
% \label{subsec:cost_model}

We adopt the cost of the QKD network components and the links te{y-cao2021-hybrid} to support the deployment of QKD. The QKD network components comprise MDI-QTxs, MDI-QRxs, LKMs, SIs, MUX/DEMUX components, and QKD and KM links. The cost of components and links can be described as follows:

%\vspace{-0.2cm}
\subsubsection{MDI-QTxs and MDI-QRxs} An MDI-QKD process requires two MDI-QTxs and one MDI-QRx, and therefore the number of MDI-QTxs $A^{f}_{\mathrm{tx}}(\cdot)$ and MDI-QRxs $A^{f}_{\mathrm{rx}}(\cdot)$ to satisfy the request $f$, can be expressed in as follows:
\beqn
		A^{f}_{\mathrm{tx}} (\tilde{\omega}, \Theta) = \sum_{i \in \mathcal{N}_{f}} \sum_{j \in \mathcal{N}_{f}} 2 P_{f}(\tilde{\omega}) \Big\lceil \frac{e_{(i,j)}} {\Theta}	\Big\rceil   \label{eq:cost-mdi-qtxs} \\
		A^{f}_{\mathrm{rx}} (\tilde{\omega}, \Theta) = \sum_{i \in \mathcal{N}_{f}} \sum_{j \in \mathcal{N}_{f}}   P_{f}(\tilde{\omega}) \Big\lceil \frac{e_{(i,j)}} {\Theta}	\Big\rceil   \label{eq:cost-mdi-qrxs}
\eeqn
$e_{(i,j)}$ is the physical distance between nodes $i$ and $j$. $\mathcal{N}_{f}$ is the set of nodes having links on the route of request $f$. 
%\vspace{-0.5cm}
\subsubsection{LKMs} For the QKD request $f$, the required number of LKMs $A^{f}_{\mathrm{km}}$ can be expressed as follows: 
\beqn
	   A^{f}_{\mathrm{km}}(\Theta) =  \sum_{i \in \mathcal{N}_{f}} \sum_{j \in \mathcal{N}_{f}} \Big\lceil \frac{e_{(i,j)}} {\Theta} + 1 \Big\rceil.   \label{eq:cost-lkms} 
\eeqn
%\vspace{-0.5cm}
\subsubsection{SIs} The required number of SIs $A^{f}_{\mathrm{si}}$ to satisfy the QKD request $f$ is expressed as follows:
\beqn
	   A^{f}_{\mathrm{si}}(\Theta) =  \sum_{i \in \mathcal{N}_{f}} \sum_{j \in \mathcal{N}_{f}} \Big\lceil \frac{e_{(i,j)}} {\Theta} - 1 \Big\rceil.   \label{eq:cost-si} 
\eeqn
%\vspace{-0.5cm}
\subsubsection{MUX/DEMUX} The required number of MUX/DEMUX pairs $A^{f}_{\mathrm{md}}$ for the QKD request $f$ is expressed as follows:
\beqn
	A^{f}_{\mathrm{md}}(\Theta) =  \sum_{i \in \mathcal{N}_{f}} \sum_{j \in \mathcal{N}_{f}} \Big\lceil \frac{e_{(i,j)}} {\Theta} \Big\rceil  +   \sum_{i \in \mathcal{N}_{f}} \sum_{j \in \mathcal{N}_{f}} \Big\lceil \frac{e_{(i,j)}} {\Theta} - 1 \Big\rceil.  \label{eq:cost-mux-demux}
\eeqn
%\vspace{-0.5cm}
\subsubsection{QKD and KM links} Three wavelengths and one wavelength are occupied by a QKD link and a KM link, respectively~\cite{a-wonfor2019-field}. The link cost for the QKD request $f$ can be expressed as follows:  
\beqn
	A^{f}_{\mathrm{ch}} ( \tilde{\omega}) =  \sum_{i \in \mathcal{N}_{f}} \sum_{j \in \mathcal{N}_{f}} ( 3 P_{f}(\tilde{\omega}) e_{(i,j)} + e_{(i,j)} ).  \label{eq:cost-qkd-km-link} 
\eeqn
Here, the total lengths of QKD links and KM links are denoted by $3P_{f}(\tilde{\omega}) e_{(i,j)}$ and $e_{(i,j)}$, respectively.

%%%%%%%%%%%%%%%%%%%%%%%%%%%%%%%%%%%%%%%%%%%%%%%%%%%%%%%%%%%

In the SP model, we define the reservation, utilization, and on-demand phases. The reservation phase is the phase that reserves QKD resources before knowing the exact QKD resources. The utilization phase is that utilizes QKD resources reserved in the reservation phase. The on-demand phase is that utilizes the on-demand QKD resources if the resources in the reservation phase are insufficient. The cost functions of QKD network components for the reservation phase are shown in (\ref{eq:reservation1}) and (\ref{eq:reservation2}), the utilization phase are shown in (\ref{eq:utilization1}) and (\ref{eq:utilization2}), and the on-demand phase are shown in (\ref{eq:on-demand1}) and (\ref{eq:on-demand2}).
\beqn
\tau (\bar{\omega}, \Theta)  =  \frac{1}{3} ( A^{f}_{\mathrm{tx}}(\bar{\omega}, \Theta) \beta^{\mathrm{r}\Theta}_{\mathrm{tx}} + A^{f}_{\mathrm{rx}}(\bar{\omega}, \Theta)  \beta^{\mathrm{r}\Theta}_{\mathrm{rx}} ) \label{eq:reservation1} \\
\lambda ( \Theta )   =   A^{f}_{\mathrm{km}}( \Theta ) \beta^{\mathrm{r}\Theta}_{\mathrm{km}} + A^{f}_{\mathrm{si}}( \Theta ) \beta^{\mathrm{r}\Theta}_{\mathrm{si}} + A^{f}_{\mathrm{md}}( \Theta ) \beta^{\mathrm{r}\Theta}_{\mathrm{md}} \label{eq:reservation2} \\
\phi (\tilde{\omega}, \Theta )  =   \frac{1}{3} ( A^{f}_{\mathrm{tx}}(\tilde{\omega}, \Theta) \beta^{\mathrm{e\Theta}}_{\mathrm{tx}} + A^{f}_{\mathrm{rx}}(\tilde{\omega}, \Theta)  \beta^{\mathrm{e}\Theta}_{\mathrm{rx}} )  \label{eq:utilization1} \\
\delta ( \Theta )  =  A^{f}_{\mathrm{km}}( \Theta ) \beta^{\mathrm{e}\Theta}_{\mathrm{km}} + A^{f}_{\mathrm{si}}( \Theta ) \beta^{\mathrm{e}\Theta}_{\mathrm{si}} + A^{f}_{\mathrm{md}}( \Theta ) \beta^{\mathrm{e}\Theta}_{\mathrm{md}} \label{eq:utilization2} \\
\psi (\tilde{\omega}, \Theta )  =  \frac{1}{3} ( A^{f}_{\mathrm{tx}}(\tilde{\omega}, \Theta) \beta^{\mathrm{o}\Theta}_{\mathrm{tx}} + A^{f}_{\mathrm{rx}}(\tilde{\omega}, \Theta)  \beta^{\mathrm{o}\Theta}_{\mathrm{rx}} ) \label{eq:on-demand1} \\ 
\xi ( \Theta )  =   A^{f}_{\mathrm{km}} ( \Theta ) \beta^{\mathrm{o}\Theta}_{\mathrm{km}} + A^{f}_{\mathrm{si}}( \Theta )  \beta^{\mathrm{o}\Theta}_{\mathrm{si}} + A^{f}_{\mathrm{md}} ( \Theta ) \beta^{\mathrm{o}\Theta}_{\mathrm{md}}  \label{eq:on-demand2}
\eeqn

\section{Optimization Formulation}
\label{sec:optimization}

\subsection{Model Description}  

The optimization of the proposed QKD service provisioning framework is based on a two-stage SP \cite{Brige1997}. The sets and decision variables in the optimization are defined as follows: 

\begin{itemize}
	\item ${\mathcal{N}}$ denotes the set of all nodes in the network.
	\item ${\mathcal{O}}_n$ denotes the set of outgoing links from node $n \in {\mathcal{N}}$.
	\item ${\mathcal{I}}_n$ denotes the set of incoming links to node $n \in {\mathcal{N}}$.
	\item ${\mathcal{F}}$ denotes the set of QKD requests in the network.
	\item $w_{i,j,f}$ is a binary variable indicating whether request $f \in {\mathcal{F}}$ will take a route with the link from node $i \in {\mathcal{N}}$ to node $j \in {\mathcal{N}}$ or not.	
	\item $x^{\mathrm{rqk}}_{i,n,f}$, $x^{\mathrm{rkm}}_{i,n,f}$, $y^{\mathrm{rqk}}_{i,n,f}$, $y^{\mathrm{rkm}}_{i,n,f}$, $z^{\mathrm{rqk}}_{i,n,f}$, and $z^{\mathrm{rkm}}_{i,n,f}$ are non-negative variables describing the reserved QKD and KM wavelengths for QKD and KM links of optical fibers, UAVs, and satellites, respectively.		
	\item $x^{\mathrm{eqk}}_{i,n,f,\omega}$, $x^{\mathrm{ekm}}_{i,n,f,\omega}$, $y^{\mathrm{eqk}}_{i,n,f,\omega}$, $y^{\mathrm{ekm}}_{i,n,f,\omega}$, $z^{\mathrm{eqk}}_{i,n,f,\omega}$, and $z^{\mathrm{ekm}}_{i,n,f,\omega}$ are non-negative variables describing the expended QKD and KM wavelengths for QKD and KM links of optical fibers, UAVs, and satellites, respectively.   
    \item $x^{\mathrm{oqk}}_{i,n,f,\omega}$, $x^{\mathrm{okm}}_{i,n,f,\omega}$, $y^{\mathrm{oqk}}_{i,n,f,\omega}$, $y^{\mathrm{okm}}_{i,n,f,\omega}$, $z^{\mathrm{oqk}}_{i,n,f,\omega}$, and $z^{\mathrm{okm}}_{i,n,f,\omega}$ are non-negative variables describing the on-demand QKD and KM wavelengths for QKD and KM links of optical fibers, UAVs, and satellites, respectively.     
\end{itemize}

We consider that the secret-key rate of request $f$ and weather conditions are uncertain. Therefore, the required secret-key rates and weather conditions are considered uncertain factors in the SP model. Let $\tilde{k}$ and $\tilde{w}$ denote the random variables of the requirement and weather condition, respectively. We can represent the composite random variable $\tilde{\omega}$ to be a scenario (i.e., $\tilde{\omega}=(\tilde{k}, \tilde{w})$). Let $\omega$ denote a scenario of request $f$. This scenario is a realization of the composite random variable $\tilde{\omega}$, and therefore we can take the value of the random variable from the set of scenarios. Let $\Psi$ and $\Omega_{f}$ denote the set of all scenarios of each requirement (i.e., a scenario space) and the set of all scenarios of request $f$, respectively. Let $\mathcal{K}$ denote the set of the secret-key rates. Let $\mathcal{W}$ denote the set of weather conditions that the satellite can provide QKD service. The set of all scenarios of each requirement can be expressed as $\Psi = {\displaystyle \prod_{f \in \mathcal{F}}} \Omega_{f} = \Omega_{1} \times \Omega_{2} \times \cdots \times \Omega_{|\mathcal{F}|}$, where $\Omega_f  = \mathcal{K} \times \mathcal{W}$. Hence, $\omega_{f}$ is the scenario space of request $f$, and $\times$ is the Cartesian product. Let $\mathbb{P}_{f}(\omega)$ be the probability that the secret-key rate of request $f$ and weather conditions will be realized.

%%%%%%%%%%%%%%% deterministic equivalent formulation version %%%%%%%%%%%%%%%%%%%%
\subsection{Optimization Formulation}
\label{subsec:def}

The SP model with the random variable $\tilde{\omega}$ can be transformed into the deterministic equivalence problem \cite{Brige1997} as expressed in (\ref{eq:def-obj}) - (\ref{eq:def-con21}). The objective function in (\ref{eq:def-obj}) is to minimize the deployment cost of the QKD services for the quantum-secured SAGIN. The decision variables are under $\omega$ (i.e., $\omega \in \Omega_f$), which means that the values of demands are available when $\omega$ is observed. 

%\begin{footnotesize}
\begin{figure*}[htb]
%\vspace{-0.1cm}
\beqn
	\min  & &  \sum_{f \in {\mathcal{F}}} \sum_{n \in {\mathcal{N}} } \sum_{i \in {\mathcal{I}}_n } \Big( A^{\mathrm{w}}_{n,f} w_{i,n,f} + \tau (\bar{\omega},\Theta_x) x^{\mathrm{rqk}}_{i,n,f} + \lambda(\Theta_x) x^{\mathrm{rkm}}_{i,n,f} + \tau(\bar{\omega},\Theta_y) y^{\mathrm{rqk}}_{i,n,f} + \lambda(\Theta_y) y^{\mathrm{rkm}}_{i,n,f} + \tau (\bar{\omega},\Theta_z) z^{\mathrm{rqk}}_{i,n,f}  \label{eq:def-obj} \nonumber \\
	& & + \lambda(\Theta_z) z^{\mathrm{rkm}}_{i,n,f} + e_{(i,n)} \big( ( x^{\mathrm{rqk}}_{i,n,f} + x^{\mathrm{rkm}}_{i,n,f} ) \beta^{\mathrm{r}\Theta_x}_{\mathrm{ch}} + ( y^{\mathrm{rqk}}_{i,n,f} + y^{\mathrm{rkm}}_{i,n,f} ) \beta^{\mathrm{r}\Theta_y}_{\mathrm{ch}} + ( z^{\mathrm{rqk}}_{i,n,f} + z^{\mathrm{rkm}}_{i,n,f} ) \beta^{\mathrm{r}\Theta_z}_{\mathrm{ch}} \big) \Big)  \nonumber \\	     	     
	& & + \sum_{f \in {\mathcal{F}}} \mathbb{P}_{f}(\omega) \sum_{n \in {\mathcal{N}} } \sum_{i \in {\mathcal{I}}_n } \Big( \phi(\omega,\Theta_x) x^{\mathrm{eqk}}_{i,n,f,\omega} + \delta(\Theta_x) x^{\mathrm{ekm}}_{i,n,f,\omega}  + \psi(\omega,\Theta_x) x^{\mathrm{oqk}}_{i,n,f,\omega} + \xi(\Theta_x) x^{\mathrm{okm}}_{i,n,f,\omega} + \phi(\omega,\Theta_y) y^{\mathrm{eqk}}_{i,n,f,\omega}  \nonumber \\
	& & + \delta(\Theta_y) y^{\mathrm{ekm}}_{i,n,f,\omega} + \psi(\omega, \Theta_y) y^{\mathrm{oqk}}_{i,n,f,\omega} + \xi(\Theta_y) y^{\mathrm{okm}}_{i,n,f,\omega} + \phi (\omega, \Theta_z) z^{\mathrm{eqk}}_{i,n,f,\omega} + \delta(\Theta_z) z^{\mathrm{ekm}}_{i,n,f,\omega} + \psi(\omega, \Theta_z) z^{\mathrm{oqk}}_{i,n,f,\omega} \nonumber \\	
	& & + \xi(\Theta_z) z^{\mathrm{okm}}_{i,n,f,\omega} + e_{(i,n)} \big( ( x^{\mathrm{eqk}}_{i,n,f,\omega} + x^{\mathrm{ekm}}_{i,n,f,\omega} ) \beta^{\mathrm{e}\Theta_x}_{\mathrm{ch}} + ( x^{\mathrm{oqk}}_{i,n,f,\omega} + x^{\mathrm{okm}}_{i,n,f,\omega} ) \beta^{\mathrm{o}\Theta_x}_{\mathrm{ch}} + ( y^{\mathrm{eqk}}_{i,n,f,\omega} + y^{\mathrm{ekm}}_{i,n,f,\omega} ) \beta^{\mathrm{e}\Theta_y}_{\mathrm{ch}} \nonumber \\	         
	& & + ( y^{\mathrm{oqk}}_{i,n,f,\omega} + y^{\mathrm{okm}}_{i,n,f,\omega} ) \beta^{\mathrm{o}\Theta_y}_{\mathrm{ch}} + ( z^{\mathrm{eqk}}_{i,n,f,\omega} + z^{\mathrm{ekm}}_{i,n,f,\omega} ) \beta^{\mathrm{e}\Theta_z}_{\mathrm{ch}}
	         + ( z^{\mathrm{oqk}}_{i,n,f,\omega} + z^{\mathrm{okm}}_{i,n,f,\omega} ) \beta^{\mathrm{o}\Theta_z}_{\mathrm{ch}} \big)
	 \Big) 
\eeqn
\vspace{-1cm}
\end{figure*}
%\end{footnotesize}

\vspace{-2cm}
\beqn
    & & \mathrm{Subject\;to:} \nonumber \\
%%%%%%%%%%%%%%%%%%%%%% routing constraint %%%%%%%%%%%%%%%%%%%%%%%%%%%%%%%%%%%%%%%%%%%%%
	& & \sum_{ j' \in {\mathcal{O}}_{S_f} }	w_{S_f,j',f} - \sum_{ i' \in {\mathcal{I}}_{S_f} }	w_{i',S_f,f} =	1,	 f	\in {\mathcal{F}}	\label{eq:def-rt-con1} \\
	& & \sum_{ i' \in {\mathcal{I}}_{D_f} } w_{i', D_f, f} - \sum_{ j' \in {\mathcal{O}}_{D_f} } w_{D_f, j', f } =	1,	 f \in {\mathcal{F}}	 \label{eq:def-rt-con2} \\
	& & \sum_{ j' \in {\mathcal{O}}_n } w_{n, j', f }	-	\sum_{i' \in {\mathcal{I}}_n } w_{i',n, f}	=	0, n \in {\mathcal{N}} \setminus S_f, D_f  \label{eq:def-rt-con3} \\
%	& & f	\in {\mathcal{F}}, n \in {\mathcal{N}} \setminus \{ S_f, D_f \}, \\
	& & \sum_{j' \in {\mathcal{O}}_n } w_{n, j', f } 	\leq	1,	 n \in {\mathcal{N}}, f \in {\mathcal{F}} \label{eq:def-rt-con4} \\ 
	& & \Delta = w_{i,n,f}, i, n \in {\mathcal{N}},  f  \in {\mathcal{F}} \label{eq:w-to-delta} \\
%%%%%%%%%%%%%%%%%%%%%%%%%%%%%%%%%%%%%%%%%%%%%%%%%%%%%%%%%%%%%%%%%%%%%%%%%%%%%%%%%%%%%%%%
& & \Delta x^{\mathrm{rqk}}_{i,n,f} \leq \mathbb{I}^{\mathrm{q}}_{i, n} \;\mbox{and}\; \Delta x^{\mathrm{rkm}}_{i,n,f} \leq \mathbb{I}^{\mathrm{k}}_{i, n} \label{eq:def-con5} \\
%	& &  i, j, n \in {\mathcal{N}},  f  \in {\mathcal{F}},  \\
	& & \Delta y^{\mathrm{rqk}}_{i,n,f} \leq I^{\mathrm{q}}_{i, n} \;\mbox{and}\; \Delta y^{\mathrm{rkm}}_{i,n,f} \leq I^{\mathrm{k}}_{i, n} \label{eq:def-con6}  \\  
%	& &  i, j, n \in {\mathcal{N}},  f  \in {\mathcal{F}}, \\
	& & \Delta z^{\mathrm{rqk}}_{i,n,f} \leq \mathbb{J}^{\mathrm{q}}_{i, n} \;\mbox{and}\; \Delta z^{\mathrm{rkm}}_{i,n,f} \leq \mathbb{J}^{\mathrm{k}}_{i, n} \label{eq:def-con7} \\ 
%	& &  i, j, n \in {\mathcal{N}},  f  \in {\mathcal{F}},  \\	
%%%%%%%%%%%%%%%%%%%%%%%%%%%%%%%%%%%%%%%%%%%%%%%%%%%%%%%%%%%%%%%%%%%%%%%%%%%%%%%%%%%%%%%%%
	& & \Delta x^{\mathrm{eqk}}_{i,n,f,\omega} \leq \Delta x^{\mathrm{rqk}}_{i,n,f}, \;\mbox{and}\; \Delta x^{\mathrm{ekm}}_{i,n,f,\omega} \leq \Delta x^{\mathrm{rkm}}_{i,n,f} \label{eq:def-con8} \\
	& & \Delta y^{\mathrm{eqk}}_{i,n,f,\omega} \leq \Delta y^{\mathrm{rqk}}_{i,n,f} \;\mbox{and}\;  \Delta y^{\mathrm{ekm}}_{i,n,f,\omega} \leq \Delta y^{\mathrm{rkm}}_{i,n,f} \label{eq:def-con9} \\
	& & \Delta z^{\mathrm{eqk}}_{i,n,f,\omega} \leq  \Delta z^{\mathrm{rqk}}_{i,n,f} \;\mbox{and}\; \Delta z^{\mathrm{ekm}}_{i,n,f,\omega} \leq  \Delta z^{\mathrm{rkm}}_{i,n,f} \label{eq:def-con10} \\  	
%%%%%%%%%%%%%%%%%%%%%%%%%%%%%%%%%%%%%%%%%%%%%%%%%%%%%%%%%%%%%%%%%%%%%%%%%%%%%%%%%%%%%%%%%%	
	& & \Delta x^{\mathrm{oqk}}_{i,n,f,\omega} \leq \mathbb{I}^{\mathrm{oq}}_{i, n} \;\mbox{and}\; \Delta x^{\mathrm{okm}}_{i,n,f,\omega} \leq \mathbb{I}^{\mathrm{ok}}_{i, n} \label{eq:def-con11}  \\
%	& &  i, j, n \in {\mathcal{N}},  f  \in {\mathcal{F}}, \forall \omega \in \Omega_{f},  \\
	& & \Delta y^{\mathrm{oqk}}_{i,n,f,\omega} \leq I^{\mathrm{oq}}_{i, n} \;\mbox{and}\; \Delta y^{\mathrm{okm}}_{i,n,f,\omega} \leq I^{\mathrm{ok}}_{i, n}  \label{eq:def-con12}  \\ 
%	& & i, j, n \in {\mathcal{N}},  f  \in {\mathcal{F}},  \forall \omega \in \Omega_{f}, \\
	& & \Delta z^{\mathrm{oqk}}_{i,n,f,\omega} \leq \mathbb{J}^{\mathrm{oq}}_{i, n} \;\mbox{and}\; \Delta z^{\mathrm{okm}}_{i,n,f,\omega} \leq \mathbb{J}^{\mathrm{ok}}_{i, n}  \label{eq:def-con13} \\
%	& & i, j, n \in {\mathcal{N}},  f  \in {\mathcal{F}}, \forall \omega \in \Omega_{f},  \\
	& & \sum_{f \in \mathcal{F}} \Delta x^{\mathrm{eqk}}_{i,n,f,\omega} \leq \mathbb{I}^{\mathrm{q}}_{i, n},\sum_{f \in \mathcal{F}} \Delta x^{\mathrm{ekm}}_{i,n,f,\omega} \leq \mathbb{I}^{\mathrm{k}}_{i, n}  \label{eq:def-con14} \\
%	& & i, j, n \in {\mathcal{N}},  \forall \omega \in \Omega_{f}, \\
	& & \sum_{f \in \mathcal{F}} \Delta y^{\mathrm{eqk}}_{i,n,f,\omega} \leq I^{\mathrm{q}}_{i, n}, \sum_{f \in \mathcal{F}} \Delta y^{\mathrm{ekm}}_{i,n,f,\omega} \leq I^{\mathrm{k}}_{i, n}  \label{eq:def-con15}  \\	
%	& & i, j, n \in {\mathcal{N}}, \forall \omega \in \Omega_{f}, \\
	& & \sum_{f \in \mathcal{F}} \Delta z^{\mathrm{eqk}}_{i,n,f,\omega} \leq \mathbb{J}^{\mathrm{q}}_{i, n},  \sum_{f \in \mathcal{F}} \Delta z^{\mathrm{ekm}}_{i,n,f,\omega} \leq \mathbb{J}^{\mathrm{k}}_{i, n}  \label{eq:def-con16} \\
%	& & i, j, n \in {\mathcal{N}}, \forall \omega \in \Omega_{f}, \\
	& & \sum_{f \in \mathcal{F}} \Delta x^{\mathrm{oqk}}_{i,n,f,\omega} \leq \mathbb{I}^{\mathrm{oq}}_{i, n}, \sum_{f \in \mathcal{F}} \Delta x^{\mathrm{okm}}_{i,n,f,\omega} \leq \mathbb{I}^{\mathrm{ok}}_{i, n} \label{eq:def-con17} \\
%	& &  i, j, n \in {\mathcal{N}}, \forall \omega \in \Omega_{f}, \\
	& & \sum_{f \in \mathcal{F}} \Delta y^{\mathrm{oqk}}_{i,n,f,\omega} \leq I^{\mathrm{oq}}_{i, n}, \sum_{f \in \mathcal{F}} \Delta y^{\mathrm{okm}}_{i,n,f,\omega} \leq I^{\mathrm{ok}}_{i, n}   \label{eq:def-con18} \\ 
%	& &  i, j, n \in {\mathcal{N}}, \forall \omega \in \Omega_{f}, \\
	& & \sum_{f \in \mathcal{F}} \Delta z^{\mathrm{oqk}}_{i,n,f,\omega} \leq \mathbb{J}^{\mathrm{oq}}_{i, n}, \sum_{f \in \mathcal{F}} \Delta z^{\mathrm{okm}}_{i,n,f,\omega} \leq \mathbb{J}^{\mathrm{ok}}_{i, n} \label{eq:def-con19} \\
%	& & i, j, n \in {\mathcal{N}}, \forall \omega \in \Omega_{f}, \\
	%%%%% selection constraint %%%%%
    & & \Delta ( x^{\mathrm{eqk}}_{i,n,f,\omega} + y^{\mathrm{eqk}}_{i,n,f,\omega} + z^{\mathrm{eqk}}_{i,n,f,\omega} ) + x^{\mathrm{oqk}}_{i,n,f,\omega}  \label{eq:def-con20} \nonumber \\
	& & + y^{\mathrm{oqk}}_{i,n,f,\omega} + z^{\mathrm{oqk}}_{i,n,f,\omega} \geq P_{f}({\omega}) W^{\mathrm{qkd}}_{f} w_{i,j,f}  
%	& & i, j, n \in {\mathcal{N}},  f  \in {\mathcal{F}}, \forall \omega \in \Omega_{f},  \\
\eeqn

\beqn
	& & \Delta ( x^{\mathrm{ekm}}_{i,n,f,\omega} + y^{\mathrm{ekm}}_{i,n,f,\omega} + z^{\mathrm{ekm}}_{i,n,f,\omega} ) + x^{\mathrm{okm}}_{i,n,f,\omega}  \label{eq:def-con21} \nonumber \\
	& & + y^{\mathrm{okm}}_{i,n,f,\omega} + z^{\mathrm{okm}}_{i,n,f,\omega} \geq P_{f}({\omega}) W^{\mathrm{kml}}_{f} w_{i,j,f}, \nonumber \\
	& &  i, j, n \in {\mathcal{N}},  f  \in {\mathcal{F}}, \forall \omega \in \Omega_{f}		
\eeqn
%\end{figure*}

%The SP model with the random variable $\tilde{\omega}$ can be transformed into the deterministic equivalence problem \cite{Brige1997} as expressed in (\ref{eq:def-obj}) - (\ref{eq:def-con21}). The objective function in (\ref{eq:def-obj}) is to minimize the deployment cost of the QKD services for the quantum-secured SAGIN. The decision variables are under $\omega$ (i.e., $\omega \in \Omega_f$) which means that the values of demands are available when $\omega$ is observed. 

The constraint in (\ref{eq:def-rt-con1}) defines that the number of outgoing routes is larger than the number of incoming routes when the node is the source node $S_f$ of the QKD request $f$. The constraint in (\ref{eq:def-rt-con2}) states that the number of incoming routes is larger than the number of outgoing routes if the node is the destination node $D_f$ of the QKD request $f$. The constraint in (\ref{eq:def-rt-con3})  defines that the number of outgoing routes must be equal to the number of incoming routes if the node is the intermediate node of the QKD request $f$. The constraint in (\ref{eq:def-rt-con4}) states that there is no loop for any QKD request, which means that there is only one outgoing route for the QKD request of any node. The constraint in (\ref{eq:w-to-delta}) defines $\Delta$ to be the same value of $w_{i,n,f}$. $\Delta$ has the same meaning of $w_{i,n,f}$.  

The constraint in (\ref{eq:def-con5}) defines that, for the optical fibers, the QKD and KM wavelengths in the reservation phase must not exceed the maximum QKD and KM wavelengths (i.e., $\mathbb{I}^{\mathrm{q}}_{i,j}$ and $\mathbb{I}^{\mathrm{k}}_{i,j}$), respectively. The constraint in (\ref{eq:def-con6}) defines that, for the UAVs, the QKD and KM wavelengths in the reservation phase must not exceed the maximum wavelengths (i.e., $I^{\mathrm{q}}_{i,j}$ and $I^{\mathrm{k}}_{i,j}$) of the QKD and KM channels, respectively. The constraint in (\ref{eq:def-con7}) defines that, for the satellites, the QKD and KM wavelengths in the reservation phase must not exceed the maximum wavelengths of the satellite QKD and KM channels (i.e., $\mathbb{J}^{\mathrm{q}}_{i,j}$ and $\mathbb{J}^{\mathrm{k}}_{i,j}$), respectively. 

For the optical fiber constraint in (\ref{eq:def-con8}), the UAV constraint in (\ref{eq:def-con9}), and the satellite constraint in (\ref{eq:def-con10}), the utilized QKD and KM wavelengths must be less than or equal to the reserved QKD and KM wavelengths. The constraint in (\ref{eq:def-con11}) defines that, for the optical fiber, the QKD and KM wavelengths in the on-demand phase are not more than the maximum QKD and KM wavelengths (i.e., $\mathbb{I}^{\mathrm{oq}}_{i,j}$ and $\mathbb{I}^{\mathrm{ok}}_{i,j}$), respectively. The constraint in (\ref{eq:def-con12}) defines that, for the UAV, the QKD and KM wavelengths in the on-demand phase are not more than the maximum QKD and KM wavelengths (i.e., $I^{\mathrm{oq}}_{i,j}$ and $I^{\mathrm{ok}}_{i,j}$), respectively. The constraint in (\ref{eq:def-con13}) defines that, for the satellite, the QKD and KM wavelengths in the on-demand phase of the satellite are not more than the maximum QKD and KM wavelengths (i.e., $\mathbb{J}^{\mathrm{oq}}_{i,j}$ and $\mathbb{J}^{\mathrm{ok}}_{i,j}$), respectively. To guarantee that the utilized QKD and KM wavelengths of all requests are not more than the maximum QKD and KM wavelengths, the constraints of optical fibers, UAVs, and satellites define in (\ref{eq:def-con14}), (\ref{eq:def-con15}), and (\ref{eq:def-con16}), respectively. Similarly, to ensure that the on-demand QKD and KM wavelengths are not more than the maximum QKD and KM wavelengths, the constraints of the optical fibers, the UAVs, and the satellites represent in (\ref{eq:def-con17}), (\ref{eq:def-con18}), and (\ref{eq:def-con19}), respectively. The constraint in (\ref{eq:def-con20}) guarantees that all the QKD wavelengths of the optical fibers, the UAVs, and the satellites have to satisfy the secret-key rate requirements (i.e., demands), and the constraint in (\ref{eq:def-con21}) ensures that all the KM wavelengths of the optical fibers, the UAVs, and the satellites have to satisfy the demands. The decision variable $w_{i,n,f}$ is a binary variable, and the rest of the decision variables are non-negative integer numbers.

\section{Performance Evaluation}
\label{sec:performance-evaluation}

%In this section, we conduce experiments with the following considerations. First, we consider that the proposed SP model can obtain the optimal deployment cost by provisioning the QKD and KM wavelengths in the reservation and on-demand phases. Second, if the source and destination nodes are not connected by optical fiber, we suggest that the SP model can produce the efficient routing to satisfy the QKD request. In addition, we show the utilization of UAVs to support the request. Finally, we show the medium utilization (i.e., optical fibers, UAVs, and satellites) when the costs of the optical fibers and UAVs increase. 

\subsection{Parameter Setting}
\label{subsec:parameter-setting}

We consider the SAGIN network topology shown in Fig.~\ref{fig:system-model} that we adopt the NSFNET topology for ground nodes \cite{y-cao2021-hybrid} and increase the additional aerial and space nodes into the topology. We set $\Theta_{x}$, $\Theta_{y}$, and $\Theta_{z}$ to be 160 km \cite{y-cao2021-hybrid}, 1 km, and 1,000 km, respectively. For the QKD and KM wavelength capacities of the optical fiber, UAV, and satellite-based QKD links in the reservation and on-demand phase, we initially set 150 and 30 for the maximum wavelengths for the QKD link between the node $i$ and the node $j$ and the maximum wavelengths for the KM link between node $i$ and node $j$, respectively. For the SP model, we consider the random number of secret-key rates and weather conditions with uniform distribution for ease of presentation. For the weather conditions, we consider two scenarios, i.e., clear and cloudy sky. The clear sky means that the secret keys can be transmitted, while the cloudy sky means that the secret keys cannot be transmitted. We consider the cost values of five QKD network components, including MDI-QTXs, MDI-QRXs, LKMs, SIs, MUX/DEMUX components, and QKD and KM links. For the reservation phase, we apply the cost values \cite{y-cao2021-hybrid}. The cost values of the components can be presented in Table \ref{table:parameter-setting1}. For the on-demand phase, the cost values of the components are set to two times the cost values of the components in the reservation phase. 

% \mathbb{I}^{\mathrm{q}}_{i, n}, I^{\mathrm{q}}_{i, n}, \mathbb{J}^{\mathrm{q}}_{i, n} 
% \mathbb{I}^{\mathrm{k}}_{i, n}, I^{\mathrm{k}}_{i, n}, \mathbb{J}^{\mathrm{k}}_{i, n}
% \mathbb{I}^{\mathrm{oq}}_{i, n}, I^{\mathrm{oq}}_{i, n}, \mathbb{J}^{\mathrm{oq}}_{i, n} 
% \mathbb{I}^{\mathrm{ok}}_{i, n}, I^{\mathrm{ok}}_{i, n}, \mathbb{J}^{\mathrm{ok}}_{i, n}

%\vspace{-0.25cm}
\begin{table}[htb] \footnotesize \caption{Reservation and utilization cost values}
\label{table:parameter-setting1}
\centering
\scalebox{0.9}{\begin{tabular}{|l|l|l|l|l|l|}\hline
{\bf Notations} & {\bf Values} & {\bf Notations} & {\bf Values} & {\bf Notations} & {\bf Values} \\ \hline
$\beta^{\mathrm{r}\Theta_x}_{\mathrm{tx}}$, $\beta^{\mathrm{e}\Theta_x}_{\mathrm{tx}}$ & 1,500 & $\beta^{\mathrm{r}\Theta_y}_{\mathrm{tx}}$, $\beta^{\mathrm{e}\Theta_y}_{\mathrm{tx}}$ & 3,000 & $\beta^{\mathrm{r}\Theta_z}_{\mathrm{tx}}$, $\beta^{\mathrm{e}\Theta_z}_{\mathrm{tx}}$ & 12,000 \\ \hline
$\beta^{\mathrm{r}\Theta_x}_{\mathrm{rx}}$, $\beta^{\mathrm{e}\Theta_x}_{\mathrm{rx}}$ & 2,250 & $\beta^{\mathrm{r}\Theta_y}_{\mathrm{rx}}$, $\beta^{\mathrm{e}\Theta_y}_{\mathrm{rx}}$ & 4,500 & $\beta^{\mathrm{r}\Theta_z}_{\mathrm{rx}}$, $\beta^{\mathrm{e}\Theta_z}_{\mathrm{rx}}$ & 22,000 \\ \hline
$\beta^{\mathrm{r}\Theta_x}_{\mathrm{km}}$, $\beta^{\mathrm{e}\Theta_x}_{\mathrm{km}}$ & 1,200 & $\beta^{\mathrm{r}\Theta_y}_{\mathrm{km}}$, $\beta^{\mathrm{e}\Theta_y}_{\mathrm{km}}$ & 2,400  & $\beta^{\mathrm{r}\Theta_z}_{\mathrm{km}}$, $\beta^{\mathrm{e}\Theta_z}_{\mathrm{km}}$ & 10,000 \\ \hline
$\beta^{\mathrm{r}\Theta_x}_{\mathrm{si}}$, $\beta^{\mathrm{e}\Theta_x}_{\mathrm{si}}$ & 150   & $\beta^{\mathrm{r}\Theta_y}_{\mathrm{si}}$, $\beta^{\mathrm{e}\Theta_y}_{\mathrm{si}}$ & 300    & $\beta^{\mathrm{r}\Theta_z}_{\mathrm{si}}$, $\beta^{\mathrm{e}\Theta_z}_{\mathrm{si}}$ & 2,000  \\ \hline
$\beta^{\mathrm{r}\Theta_x}_{\mathrm{md}}$, $\beta^{\mathrm{e}\Theta_x}_{\mathrm{md}}$ & 300   & $\beta^{\mathrm{r}\Theta_y}_{\mathrm{md}}$, $\beta^{\mathrm{e}\Theta_y}_{\mathrm{md}}$ & 600    & $\beta^{\mathrm{r}\Theta_z}_{\mathrm{md}}$, $\beta^{\mathrm{e}\Theta_z}_{\mathrm{md}}$ & 1,000  \\ \hline
$\beta^{\mathrm{r}\Theta_x}_{\mathrm{ch}}$, $\beta^{\mathrm{e}\Theta_x}_{\mathrm{ch}}$ & 1     & $\beta^{\mathrm{r}\Theta_y}_{\mathrm{ch}}$, $\beta^{\mathrm{e}\Theta_y}_{\mathrm{ch}}$ & 2      & $\beta^{\mathrm{r}\Theta_z}_{\mathrm{ch}}$, $\beta^{\mathrm{e}\Theta_z}_{\mathrm{ch}}$ & 20     \\ \hline
\end{tabular}}
\end{table}
%\vspace{-0.25cm}

%%%%%%%%%%%%%%%%%%%%%% on-demand cost %%%%%%%%%%%%%%%%%%%%%%%%%%%%%
% \begin{table}[htb] \footnotesize \caption{On-demand cost values}
% \label{table:parameter-setting2}
% \centering
% \scalebox{0.9}{\begin{tabular}{|l|l|l|l|l|l|}\hline
% {\bf Notations} & {\bf Values(\$)} & {\bf Notations} & {\bf Values(\$)} & {\bf Notations} & {\bf Values(\$)} \\ \hline
% $\beta^{\mathrm{o}\Theta_x}_{\mathrm{tx}}$ & 6,000 & $\beta^{\mathrm{o}\Theta_y}_{\mathrm{tx}}$ & 18,000 & $\beta^{\mathrm{o}\Theta_z}_{\mathrm{tx}}$ & 20,000 \\ \hline
% $\beta^{\mathrm{o}\Theta_x}_{\mathrm{rx}}$ & 9,000 & $\beta^{\mathrm{o}\Theta_y}_{\mathrm{rx}}$ & 27,000 & $\beta^{\mathrm{o}\Theta_z}_{\mathrm{rx}}$ & 30,000 \\ \hline
% $\beta^{\mathrm{o}\Theta_x}_{\mathrm{km}}$ & 3,000 & $\beta^{\mathrm{o}\Theta_y}_{\mathrm{km}}$ & 9,000  & $\beta^{\mathrm{o}\Theta_z}_{\mathrm{km}}$ & 18,000\\ \hline
% $\beta^{\mathrm{o}\Theta_x}_{\mathrm{si}}$ & 500   & $\beta^{\mathrm{o}\Theta_y}_{\mathrm{si}}$ & 1,500  & $\beta^{\mathrm{o}\Theta_z}_{\mathrm{si}}$ & 10,000 \\ \hline
% $\beta^{\mathrm{o}\Theta_x}_{\mathrm{md}}$ & 900   & $\beta^{\mathrm{o}\Theta_y}_{\mathrm{md}}$ & 2,700  & $\beta^{\mathrm{o}\Theta_z}_{\mathrm{md}}$ & 9,000\\ \hline
% $\beta^{\mathrm{o}\Theta_x}_{\mathrm{ch}}$ & 4     & $\beta^{\mathrm{o}\Theta_y}_{\mathrm{ch}}$ & 12     & $\beta^{\mathrm{o}\Theta_z}_{\mathrm{ch}}$ & 100 \\ \hline
% \end{tabular}}
% \end{table}

\subsection{Numerical Results}
\label{subsubsec:numerical-results}

\subsubsection{Cost Structure Analysis}
\label{subsubsec:cost-stucture-analysis}

\begin{figure}[htb]
\vspace{-0.2cm}
 \centering
  \subfloat[The solution of SP model]{\label{fig:qkd-wavelength-cost}\includegraphics[width=0.25\textwidth]{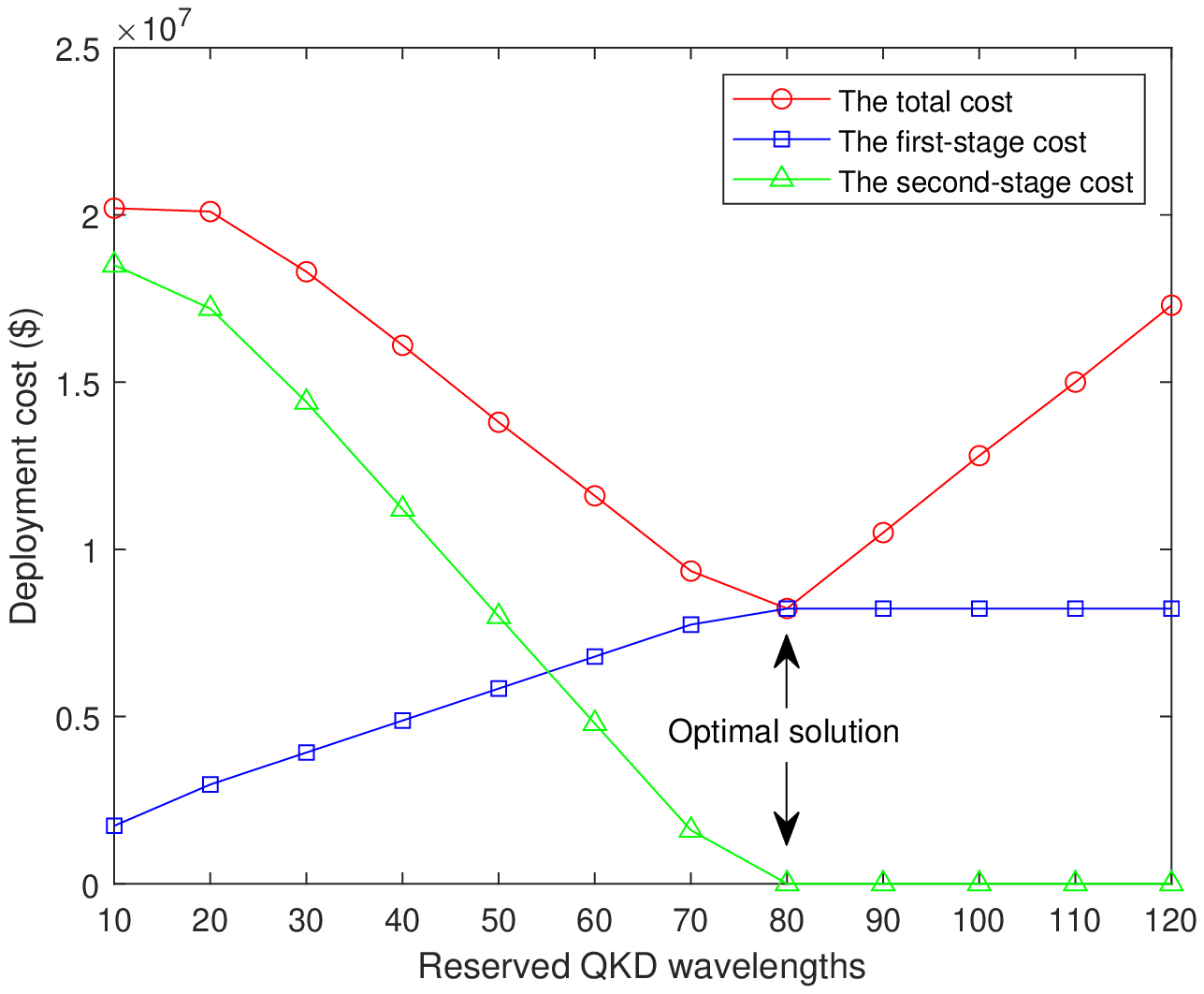}} 
  \subfloat[The QKD and KM wavelengths]{\label{fig:qkd-km-secret-key-rates}\includegraphics[width=0.25\textwidth]{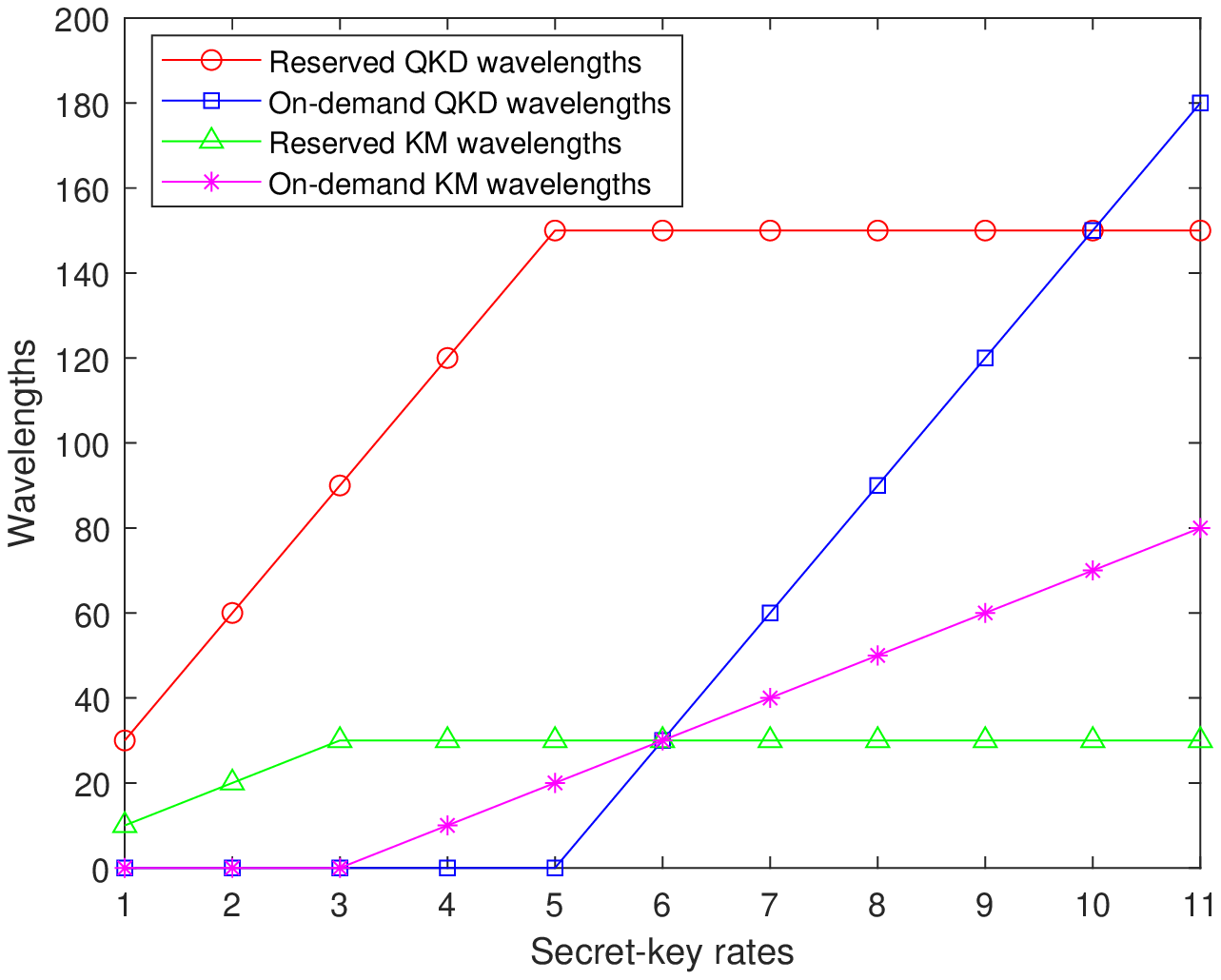}} \\[-2ex]
  \subfloat[The deployment cost]{\label{fig:km-wavelength-cost}\includegraphics[width=0.25\textwidth]{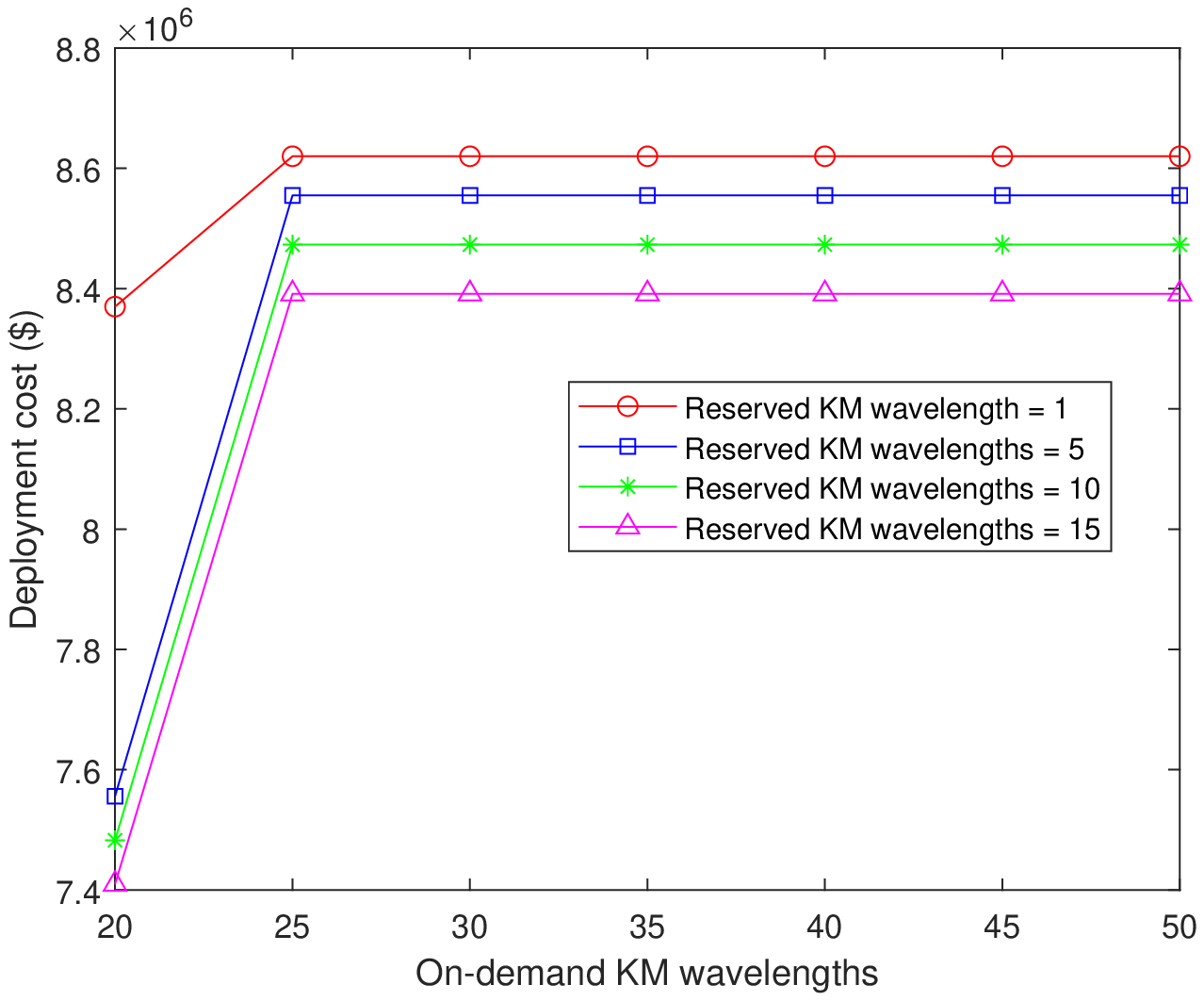}}  
  \subfloat[The deployment cost]{\label{fig:km-wavelength-cost}\includegraphics[width=0.25\textwidth]{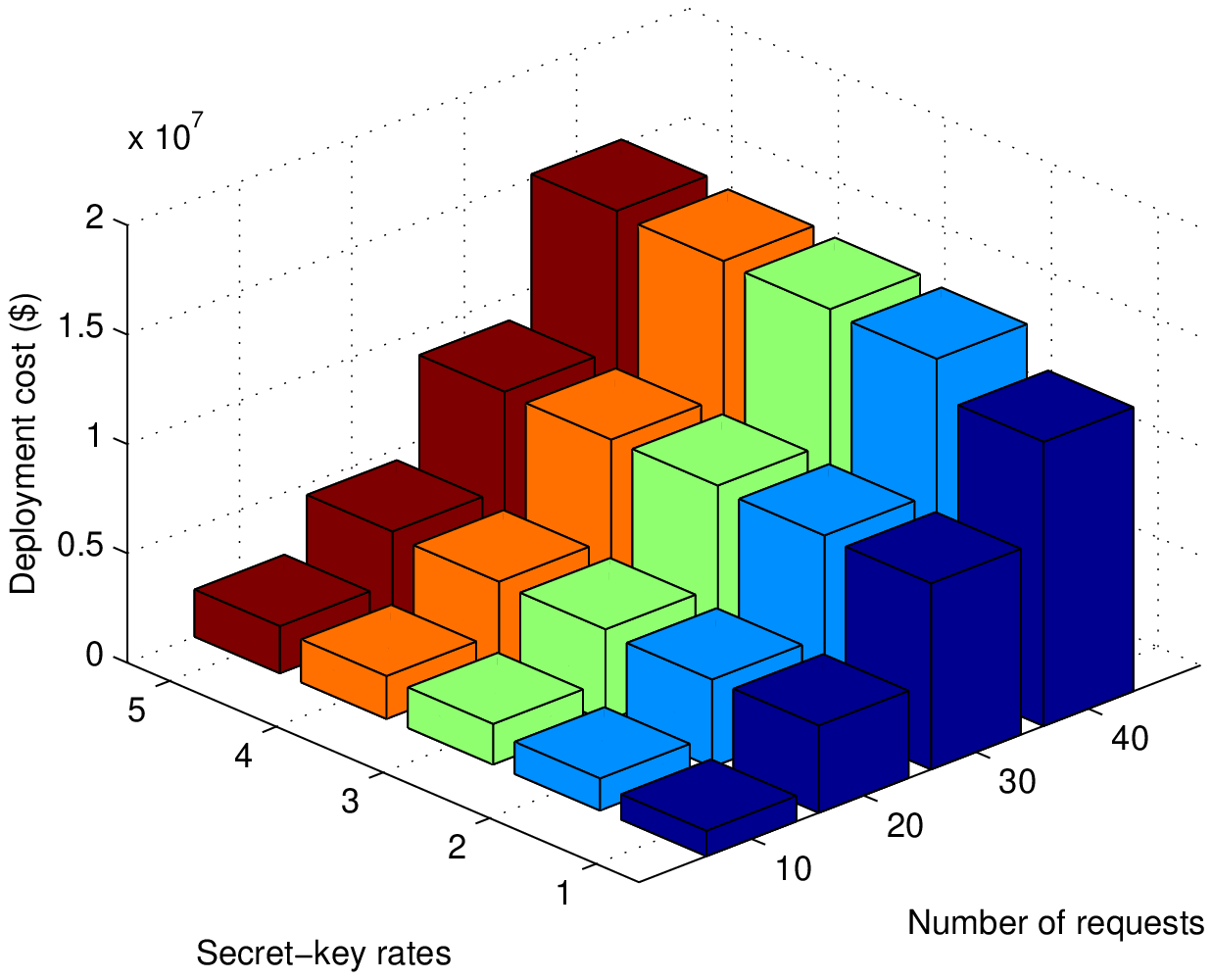}}
 \caption{(a) The optimal solution under different reserved QKD wavelengths, (b) The QKD and KM wavelengths in the reservation and on-demand phases under different secret-key rates, (c) The deployment cost of reserved KM wavelengths under different on-demand KM wavelengths, and (d) The deployment cost under different secret-key rates and requests.}
 \label{fig:solution-cost-sp}
% \vspace{-0.3cm}
\end{figure}  

As illustrated in Fig.~\ref{fig:solution-cost-sp}(a), we examine the ability of the SP model to obtain the optimal solution. In the first stage, the reserved QKD wavelengths are varied while the reserved KM wavelengths are constant. Then, we present the optimal solution and the effect of reserved QKD wavelengths on the solution. In Fig.~\ref{fig:solution-cost-sp}(a), the first-stage cost increases when the reserved QKD wavelengths rise. Nevertheless, the second-stage cost reduces dramatically when the secret-key rates are observed. This is due to the fact that the cheaper QKD wavelengths in the reservation phase (i.e., the first stage) are utilized instead of the QKD wavelengths in the on-demand phase (i.e., the second stage). As a result, the optimal solution can be achieved at the reserved QKD wavelength 80 while the second-stage cost is 0. This is because the reserved QKD wavelengths satisfy the secret-key rates (i.e., demands), and the on-demand QKD wavelengths in the second stage are not utilized. After the QKD wavelengths increase to 80, the total cost and the first-stage cost increase since there is the penalty cost for the excess of reserved QKD wavelengths to be charged. For Fig.~\ref{fig:solution-cost-sp}(a), we can see that the total cost depends on the over- and under-provision of QKD wavelengths. 

Fig.~\ref{fig:solution-cost-sp}(b) shows the QKD and KM wavelengths in the reservation and on-demand phases under different secret-key rates. In the reservation phase, the QKD and KM wavelengths increase until the secret-key rates reach 5 and 3 kbps, respectively. This is because the reserved QKD and KM wavelengths are fully utilized at the secret-key rates of 5 and 3 kbps, respectively. After the points, both QKD and KM wavelengths are constant due to the limited capacities of QKD and KM wavelengths in the reservation phase. As a result, to satisfy the high secret-key rates, the QKD and KM wavelengths in the on-demand phase are more utilized at 5 and 3 kbps, respectively.

Figure~\ref{fig:solution-cost-sp}(c) shows the deployment cost under different reserved KM wavelengths. We can observe that the minimum cost is obtained when more KM wavelengths are reserved. The reason is that the reserved KM wavelengths are utilized to satisfy the secret-key rates before the on-demand KM wavelengths. The on-demand KM wavelengths compensate for the secret-key rates that cannot be satisfied if there are not enough reserved KM wavelengths. Figure~\ref{fig:solution-cost-sp}(d) shows the deployment cost under different secret-key rates and requests. The deployment cost increases when the secret-key rates and requests increase. Therefore, we can explain that the secret-key rates and requests have a significant effect on the deployment cost.

\subsubsection{Performance Evaluation Under Various Parameters}
\label{subsubsec:performance-ev-various-parameters}

\vspace{-0.5cm}
\begin{figure}[htb]
 \centering
 \subfloat[The routing]{\label{fig:routing-nsfnet}\includegraphics[width=0.25\textwidth]{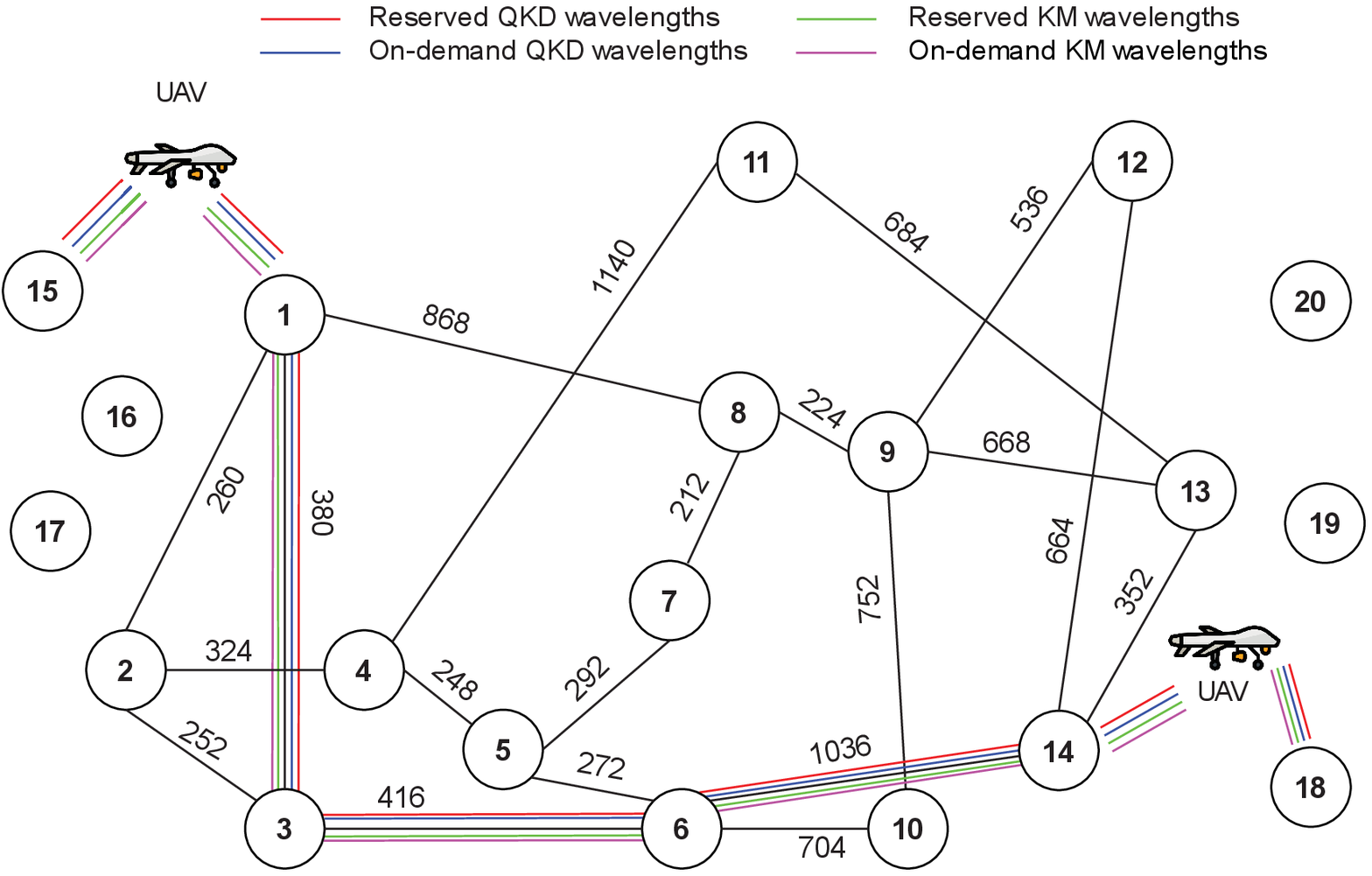}}
 \subfloat[The QKD and KM wavelengths]{\label{fig:qkd-km-under-secret-key-rates-fiber}\includegraphics[width=0.25\textwidth]{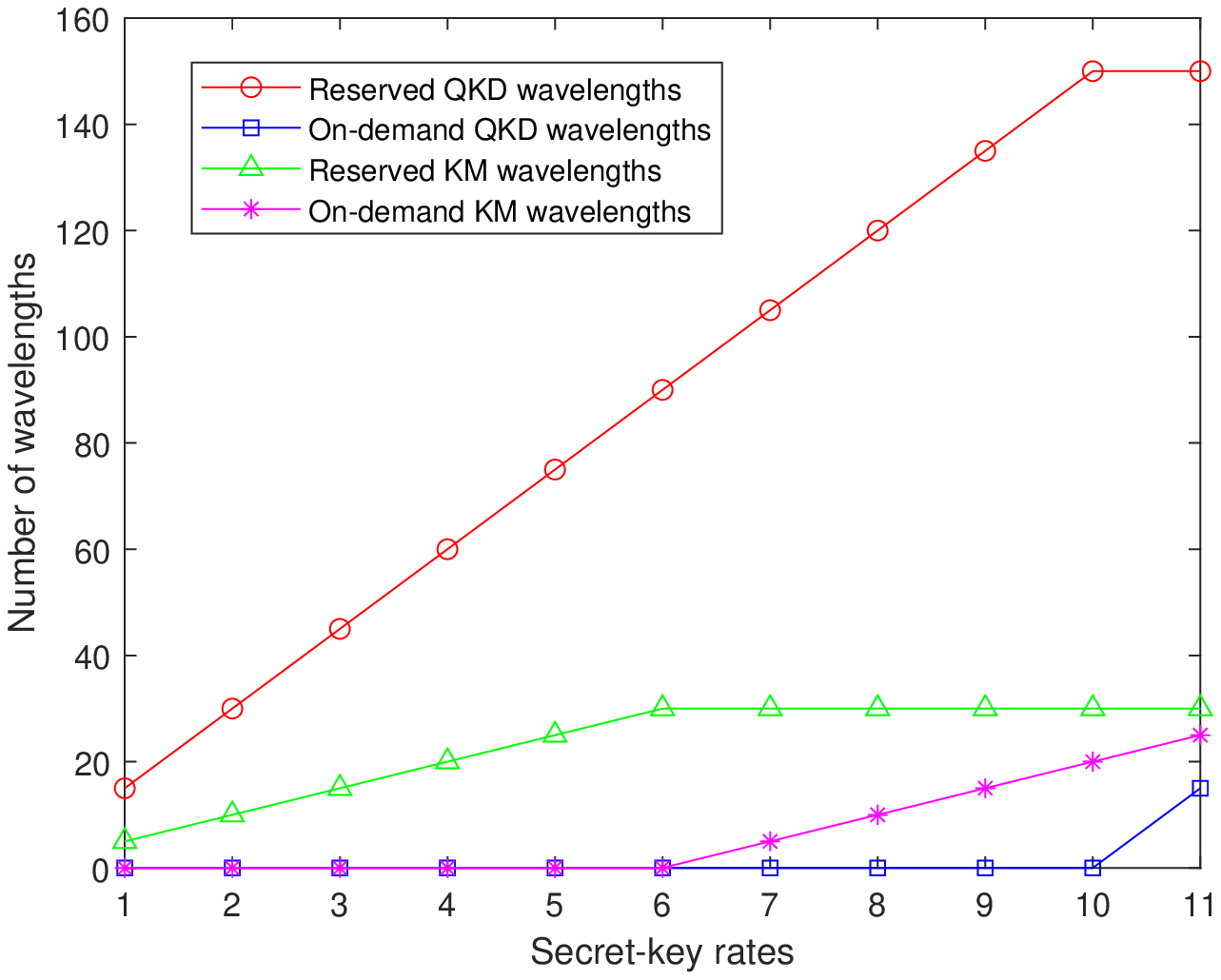}} \\[-2ex]
  \subfloat[The trajectory of the UAV]{\label{fig:qkd-km-under-secret-key-rates-uav}\includegraphics[width=0.23\textwidth]{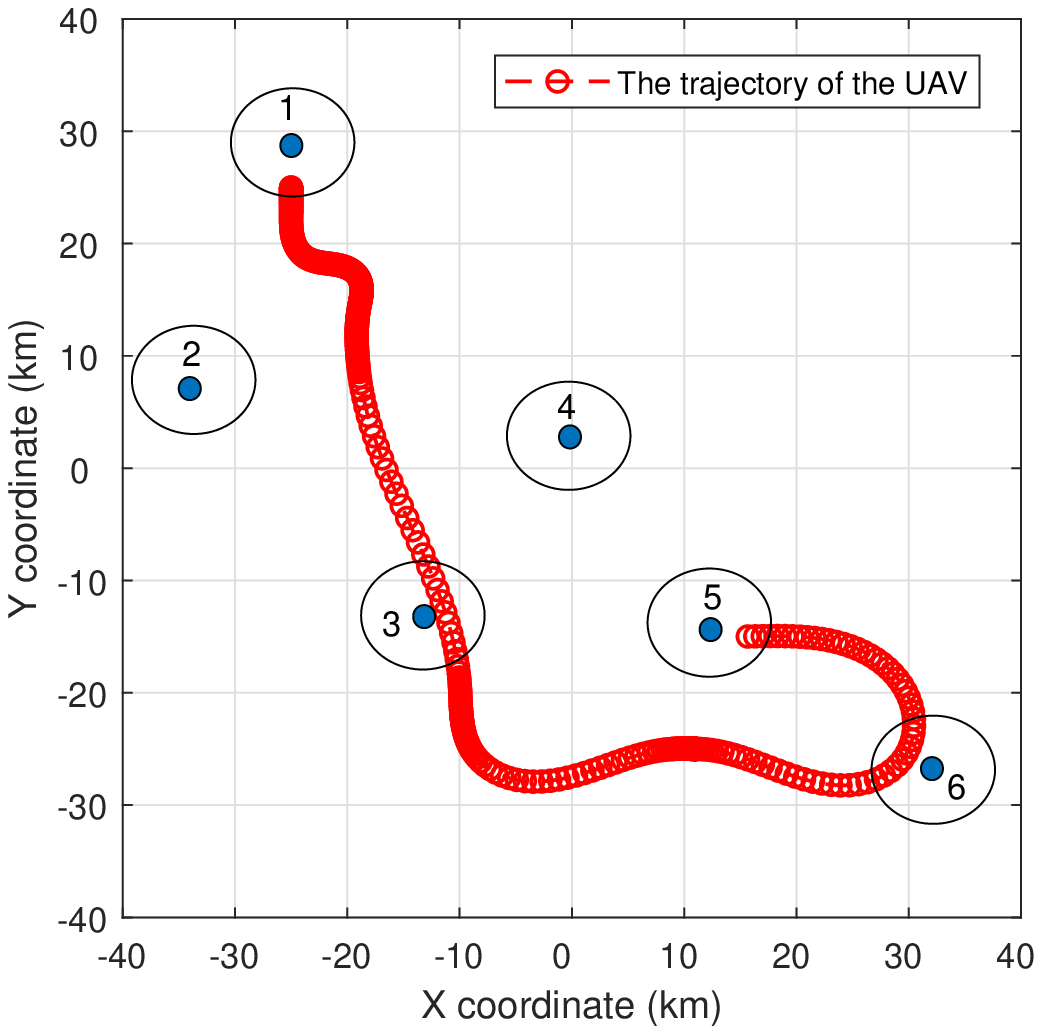}}
  \subfloat[The transition of medium utilization]{\label{fig:qkd-km-under-secret-key-rates-uav}\includegraphics[width=0.25\textwidth]{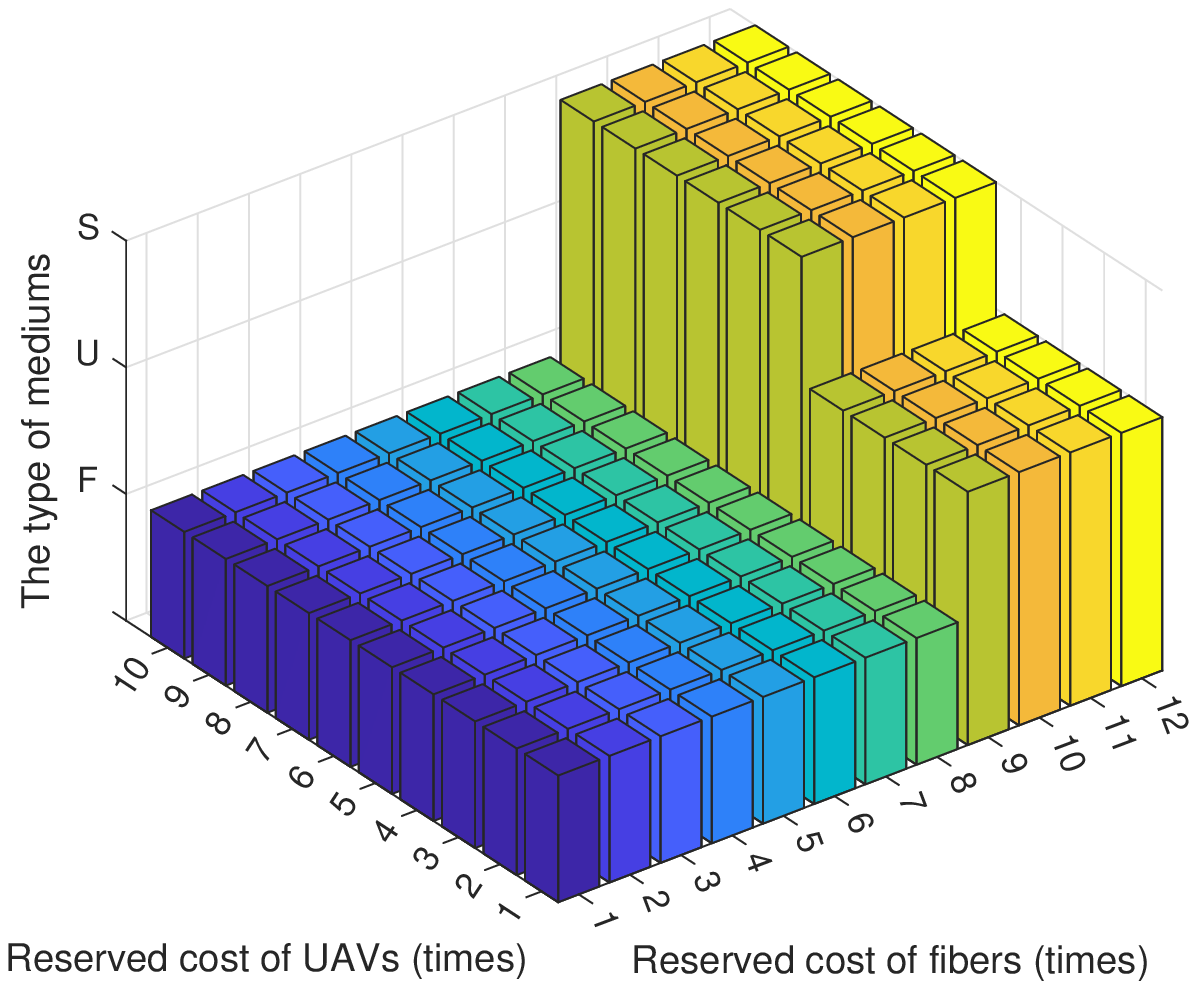}}
 \caption{(a) The routing, (b) The QKD and KM wavelengths in reservation and on-demand phases of optical fibers and UAVs under different secret-key rates, (c) The trajectory of the UAV, and (d) The transition of the medium utilization.}
 \label{fig:routing-qkd-km-fiber-uav}
%\vspace{-0.3cm}
\end{figure} 

Figure~\ref{fig:routing-qkd-km-fiber-uav}(a) shows the solution of the SP model that satisfies the request (i.e., from source node 15 to destination node 18). In the solution, the QKD and KM wavelengths in the reservation and on-demand phases are allocated in the optical fibers and UAVs. To achieve the optimal cost, both QKD and KM wavelengths in the reservation and on-demand phases are utilized along the route. Figure~\ref{fig:routing-qkd-km-fiber-uav}(b) shows the QKD and KM wavelengths provided for the request in the reservation and on-demand phases under different secret-key rates. It is clear that the on-demand QKD and KM wavelengths are utilized if the reserved QKD and KM wavelengths are fully utilized. 

We investigate a case study in there are only the UAV and satellite provisions for the request (i.e., from the source node 1 to the destination node 5) to show how the solution of the SP model is achieved. In Fig.~\ref{fig:routing-qkd-km-fiber-uav}(c), the trajectory of the UAV (i.e., solution) to satisfy the request can be obtained by the SP model. In particular, the trajectory of the UAV starts from node 1 and stops at node 5 by flying along the path that is consisted of nodes 1, 3, 6, and 5. The QKD and KM wavelengths are reserved for the trajectory. Therefore, with this case study, the SP model can efficiently produce the optimal solution.     

We consider a situation where optical fibers, UAVs, and satellites have only QKD services in the reservation phase. Hence, we increase the cost of five components of optical fibers and UAVs in the reservation phase to show the adaptation of the medium utilization (i.e., optical fibers, UAVs, and satellites). Figure~\ref{fig:routing-qkd-km-fiber-uav}(d) shows the adaptation of the medium utilization under different costs of optical fibers and UAVs in the reservation phase. In Fig.~\ref{fig:routing-qkd-km-fiber-uav}(d), F, U, and S on the axis of type of mediums are defined as the utilization of optical fibers, UAVs, and satellites, respectively. Clearly, with the high cost of five components of optical fibers and UAVs, the UAVs (i.e., U) and satellites (i.e., S) will be utilized. 

%For example, the satellite is utilized when the reservation costs of optical fibers and UAVs increase 9 times and 5 times of their original costs, respectively. The satellite utilization is marked with number 3 shown in Fig.~\ref{fig:routing-qkd-km-fiber-uav}(d). The reason is that, to achieve the optimal cost, the SP model makes a decision using the satellites instead of optical fibers and UAVs.  

\section{Conclusion}
\label{sec:conclusion}
In this paper, we have studied the quantum-secured SAGIN system and resource allocation model for QKD over SAGIN. We have proposed the resource allocation for the QKD over SAGIN, where the optical fiber-based QKD, UAV-based QKD, and satellite-based QKD services are provided to secure communications. We have formulated the joint resource allocation for the QKD nodes and links as the stochastic programming model to minimize the deployment cost under the uncertainties of the secret-key rate requirements and the weather conditions. The experimental results have clearly shown that the optimal deployment cost has been obtained by the proposed model.

\end{document}